# Initiation stage of nanosecond breakdown in liquid

Mikhail Pekker[1,1], Yohan Seepersad[1, 2], Mikhail Shneider[3], Alexander Fridman[1, 4], Danil Dobrynin[1]

*1 - A.J. Drexel Plasma Institute, Camden, NJ, 08103, USA*
*2 - Department of Electrical and Computer Engineering, Drexel University, Philadelphia, Pa, 19104, USA*
*3 - Department of Mechanical and Aerospace Engineering, Princeton University, Princeton, NJ, 08544, USA*
*4 - Department of Mechanics and Mechanical Engineering, Drexel University, Philadelphia, Pa, 19104, USA*

## Abstract

In this paper, based on a theoretical model [1], it has been shown experimentally that the initial stage of development of a nanosecond breakdown in liquids is associated with the appearance of discontinuities in the liquid (cavitation) under the influence of electrostriction forces. Comparison of experimentally measured area dimensions and its temporal development were found to be in a good agreement with the theoretical calculations.

## Introduction

The experimental study of nanosecond breakdown in the liquid phase has recently become possible with the development of stable and reliable high voltage nanosecond pulsed power sources [8-10]. Typically in the vicinity of the electrode, bubbles are generated due to ohmic heating via a standard breakdown mechanism [11], however, this mechanism does not work in a few nanoseconds. Other mechanisms proposed in [12], [13] also could not explain the breakdown with pulses of such a short time.

The breakdown mechanism in water, based on the electrostriction effect, was proposed in [2]. It is well known that in a non-uniform electric field, volumetric ponderomotive forces pull a dielectric into the region with strongest field [14], [15]. However, if the voltage rise time is very steep, the fluid does not have enough time to come into motion due to inertia. Consequently, the ponderomotive forces cause significant electrostrictive tensile stress in the dielectric liquid, which can lead to a disruption of the continuity of liquid (creating nano-pores), similar to those observed in [16]-[18] and many other experiments.

For the formation of discontinuities, it is necessary that the negative pressure in the fluid reach values on the order of 10-30 MPa [16]. In this case, density fluctuations, which always exist due to thermal motion, lead to the formation of growing cavitation bubbles (voids). However, in the case of a long rise time of the high voltage pulse, pondermotive forces are able to cause movement and arising hydrostatic pressure can compensate the negative electrostrictive pressure. In other words the sum of the electrostrictive and hydrostatic pressures cannot reach the threshold for the formation of growing cavitation bubbles. In [1] it is shown how quickly the voltage on the tip of electrode should grow, so that the total pressure near the electrode reaches a critical value. The numerical model [1] was also used for modeling the experiments [5].

As it was noted in [2], voids in the liquid dielectric can be a source of secondary electrons. This is a necessary condition for the development of the discharge. If the cavities are large enough, the electrons in them can gain sufficient energy to ionize the molecules of water on the opposite side of the cavity. If the tip electrode is positively charged, the electrons quickly neutralized on the electrode, and the remaining positive charge becomes a non-uniform electric field source, which leads to creating new voids.

[1] pekkerm@gmail.com

In present work, on the basis of comparing experimental results with numerical calculations, it was shown, that electrostriction forces really lead to creating nanopores, which determine the discharge formation in [3-7].

## Numerical model

According to [1], the system of time-varying equations for a compressible fluid (water) in a strong electric field has the form:

$$\frac{\partial \rho}{\partial t} + \nabla(\rho \vec{u}) = 0 \tag{1}$$

$$\rho\left(\frac{\partial \vec{u}}{\partial t} + (\vec{u} \cdot \nabla)\vec{u}\right) = -\nabla p + \frac{\varepsilon_0}{2}\left(\frac{\partial \varepsilon}{\partial \rho}\rho\right)\nabla E^2 \tag{2}$$

$$p = (p_0 + B)\left(\frac{\rho}{\rho_0}\right)^{\gamma} - B \tag{3}$$

$$\rho_0 = 1000 \ \text{kg/m}^3, \quad p_0 = 10^5 \ \text{Pa}, \quad B = 3.07 \cdot 10^8 \ \text{Pa}, \ \gamma = 7.5$$

Here, $\rho$ is the density, $\vec{u}$ is the velocity, $p$ is the hydrostatic pressure. Equation (1) is the continuity equation, (2) is the momentum equation, and (3) is the Tait equation of state, which relates the pressure to the density of water [19], [20]. The second term in the right side of (2) corresponds to electrostrictive forces in a non-uniform electric field associated with the tensions within the dielectric, $\varepsilon_0$ is the vacuum dielectric permittivity, $\varepsilon$ is the relative dielectric constant of the medium. In polar dielectrics: $\frac{\partial \varepsilon}{\partial \rho}\rho = \alpha\varepsilon$, for water $\alpha \approx 1.5$ [21],[22].

In equation (2) friction is neglected because a few tens of nanoseconds is not enough time for boundary layer formation (estimations are shown in [1]).

**Boundary conditions.** We assumed the no-slip condition (the fluid velocity at the electrode goes to zero) and the continuity of the fluxes of the density and momentum on both the electrode and on the boundaries of the computational domain. The system (1-3) was numerically calculated in prolate spheroidal coordinates [1, 22].

Since $\frac{\varepsilon_0}{2}\left(\frac{\partial \varepsilon}{\partial \rho}\rho\right)$ is constant, we can rewrite the right part of (2) as

$$-\nabla p + \frac{\varepsilon_0}{2}\left(\frac{\partial \varepsilon}{\partial \rho}\rho\right)\nabla E^2 = -\nabla\left(p - \frac{1}{2}\alpha\varepsilon_0\varepsilon E^2\right) = -\nabla p_{total} \tag{4}$$

This means that the total pressure $p_{total} = p - \frac{1}{2}\alpha\varepsilon_0\varepsilon E^2$ acting on the dielectric liquid is a sum of hydrostatic $p$ and electrostriction-related $p_E = -\frac{1}{2}\alpha\varepsilon_0\varepsilon E^2$ pressures.

## Experimental setup

In this paper, we omit the details of the experiment, since they are the same as in [5] and [7] and focus only on the idea of the method of detection of voids proposed in [2]. Since the sharp edge of the region with voids leads to the scattering of light (the opalescence [23]), subtracting the illumination on the screen with the voltage applied (Fig. 1b) by the illumination on the screen with no voltage (Fig. 1a), we can easily define the cavitation boundary. The Schlieren method, which is a modification of the shadow [23], was used in [5], [7]. In Figure 2, the profiles of the voltage supplied to the electrode are presented [7]. Figures 3 and 4 present Schlieren images (image size 340 x 230 μm). The "dark" area in the vicinity of the electrode shows cavitation. It can be observed that there is less cavitation for a voltage $V_0$=12.2 kV than at $V_0$=20 kV, and at $V_0$=10 kV cavitation is totally absent.

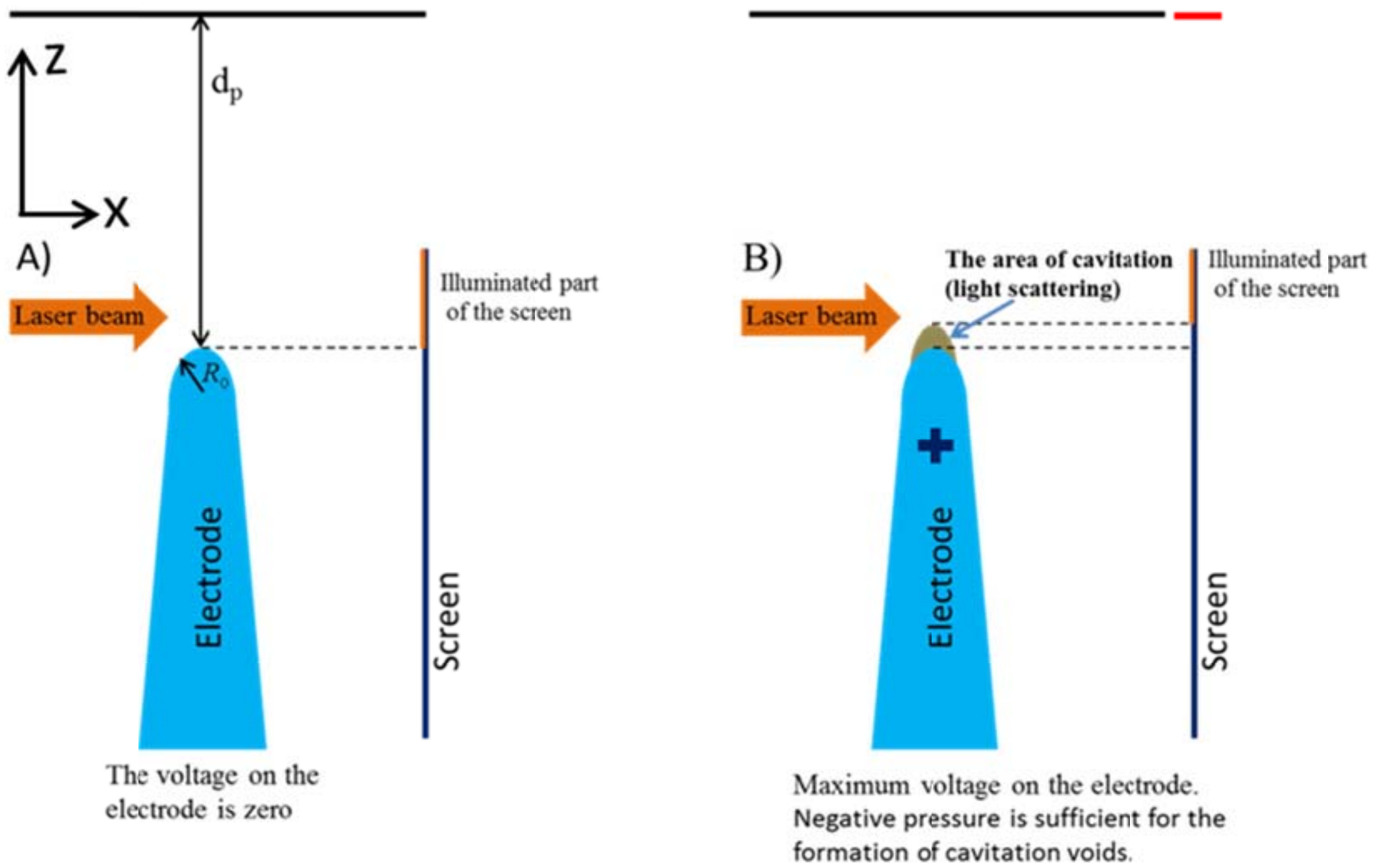

Figure 1. Optical observation of the cavitation region. The radius of curvature of the electrode is $R_0 = 35\mu m$. The distance between the needle tip and the flat electrode is $d_p = 1.5$mm. A) The voltage between the electrodes is zero, B) The voltage between the electrodes is maximum.

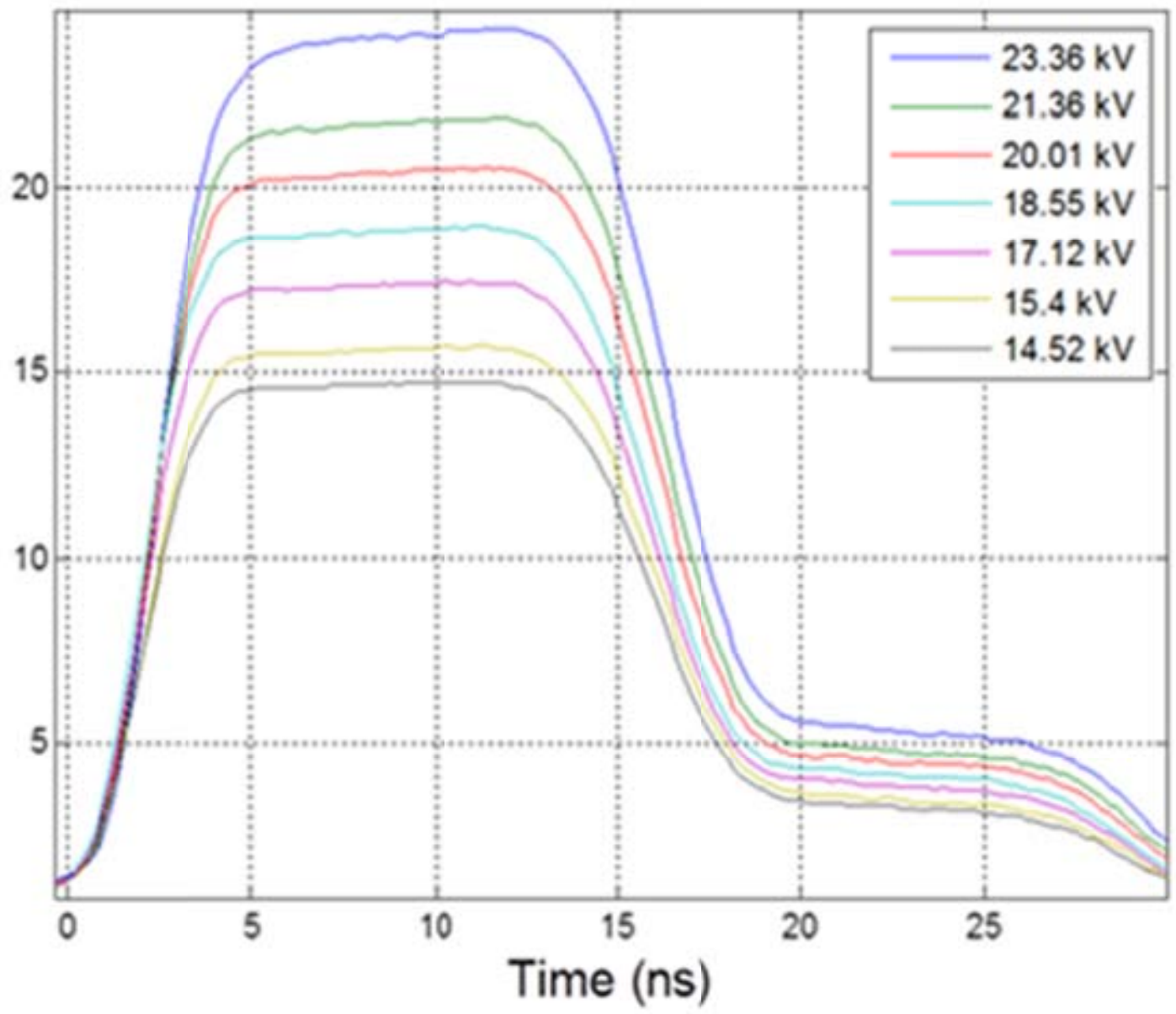

Figure 2. Time dependence of voltage on the electrode.

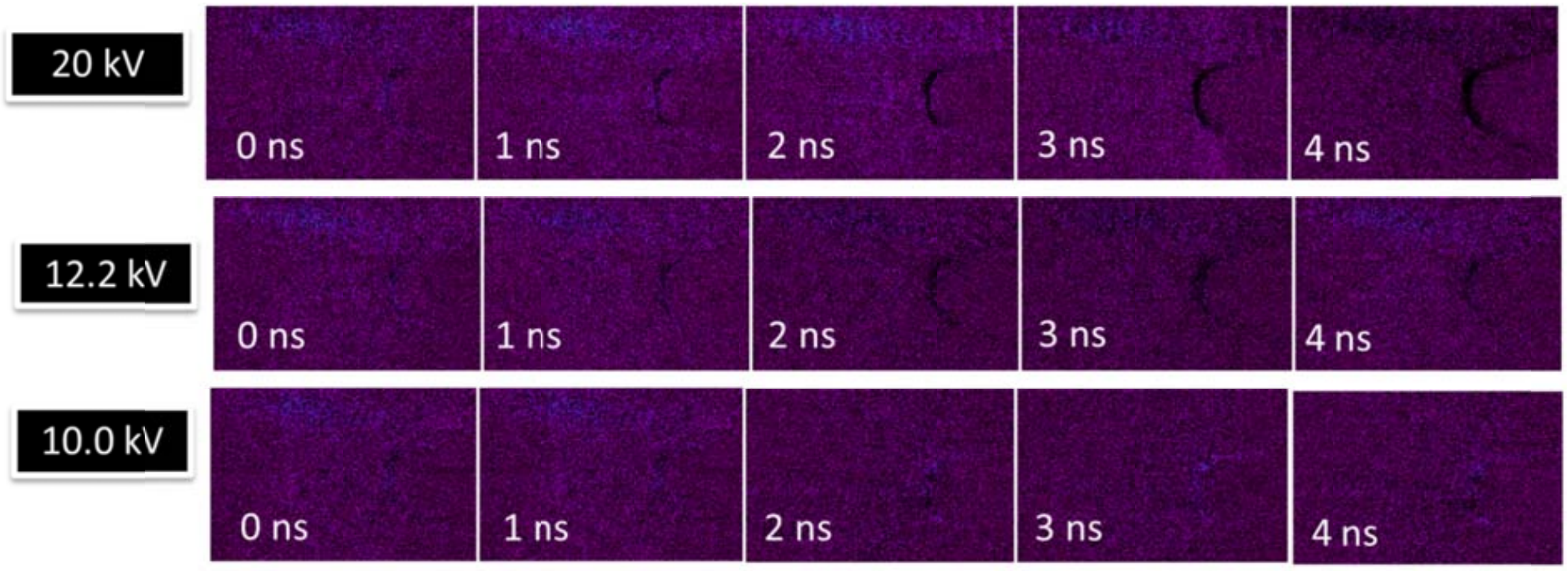

Figure 3. Results of measurement. 0ns corresponds to light accumulation from time $t_0 = 0$ ns till $t_1 = 1$ ns, 1ns – $t_0 = 1$ ns, $t_1 = 2$ ns, 2ns – $t_0 = 2$ ns, $t_1 = 3$ ns, 3ns – $t_0 = 3$ ns, $t_1 = 4$ ns, 4ns – $t_0 = 4$ ns, $t_1 = 5$ ns.

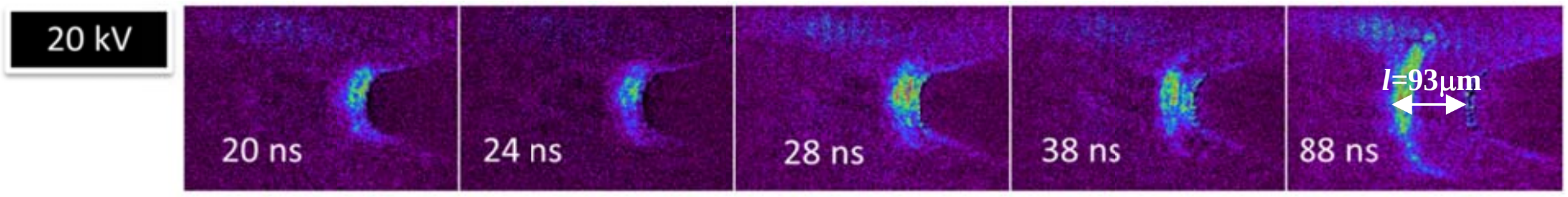

Figure 4. The results of measurements, as shown in Fig. 3, at greater time moments. $l \approx 111 \mu m$

**1. Numerical simulation.**

The results of calculations fulfilled on the basis of equations (1-3) for the parameters similar to the experiment described in this paper are presented in Figure 5. The radius of curvature of the needle electrode was $R_0 = 35\mu m$, the amplitude of the voltage on the electrode was $V_0$= 20kV, 12.2kV, 10kV, the time dependence of the voltage was varied in accordance with the Fig. 2: in time interval 0-4ns, voltage grew linearly, in 4-14ns was constant, in 14-18ns dropped linearly, in 18-150 ns, the voltage was zero.

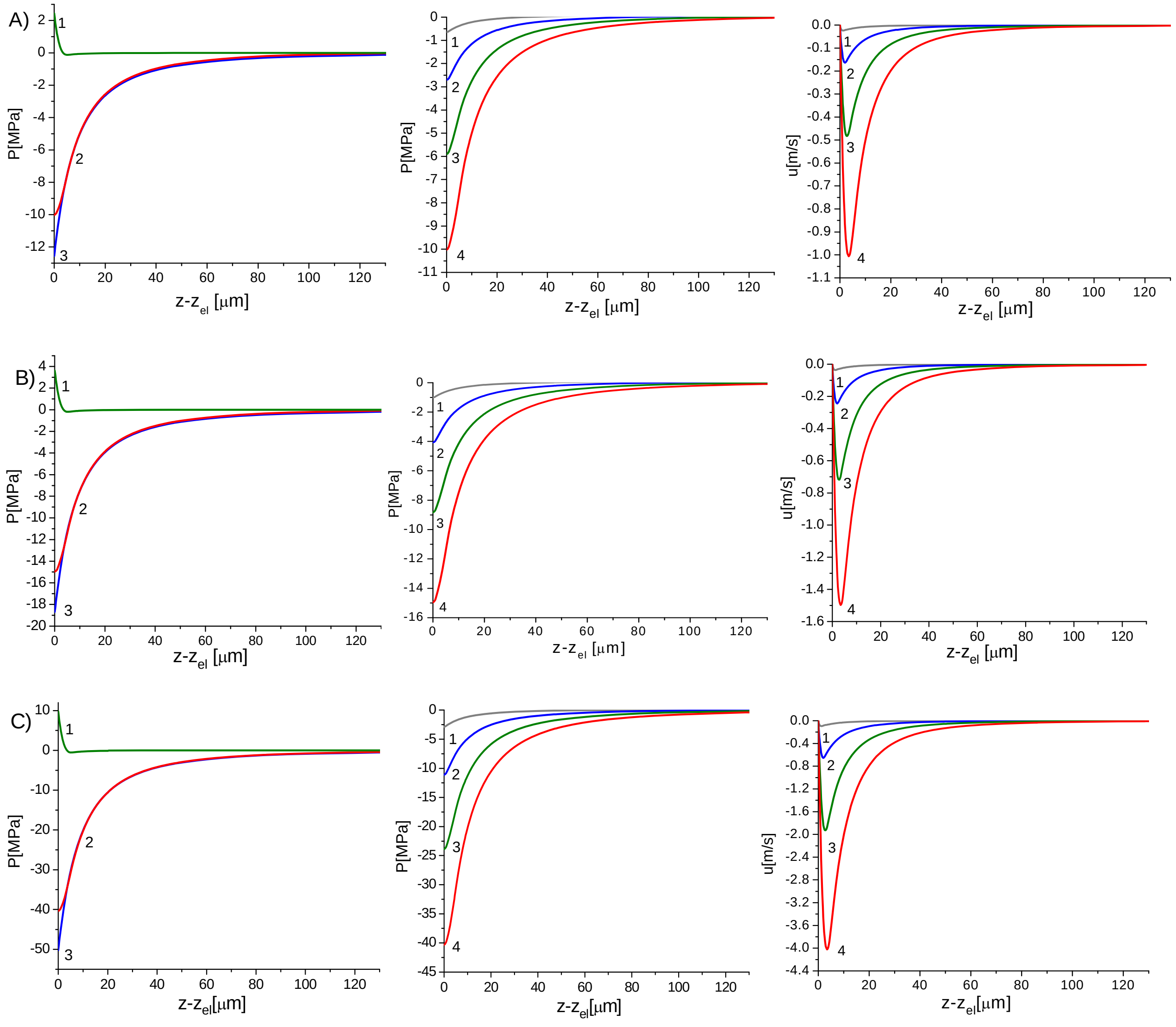

Figure. 5 Distributions of the pressure and velocity along axes of the electrode. Figures A) correspond to $V_0$ 10kv, B) – 12.2 kV, C) – 12.2 kV. In the first column of figures results of calculation at time t=4ns are shown. Curves 1 correspond to hydrostatic pressure $p$, 2 is the total pressure $p_{total}$, 3 is the electrostriction pressure $p_E$. In columns 2 and 3 dependencies total pressure and velocity $u$ 3 are shown at different moment of time. Curves 1 – t=1ns, 2 – 2ns, 3 – 3ns, 4 – 4ns. $z_{el}$ - coordinate of the edge of sharp electrode.

It can be seen that due to the inertia, velocity of fluid flows are small, and does not exceed 1-5 m/s (third column on Fig. 5). This means that in the experiments described above, the liquid does not have time to reach the electrode and compensate the negative pressure related to the electrostrictive forces $p << |p_E|$, (see the first column1 in Fig.5). Thus, we can conclude that at the stage of growth amplitude of the voltage, the refractive index is small and can be neglected (see first column on Fig. 5). The corresponding estimates confirm this statement. The “dark” region observed in Figure 3 indicates the presence of voids (opalescence effect [23]). It should be noted that a threshold exists as illustrated by the absence of "dark" area that occurs at voltages less than 10keV.

Let us estimate the size of the cavitation area. Since the 12kV voltage amplitude corresponds to the maximum negative pressure equal to 12Mpa, and the 10 kV – 10Mpa, therefore, using threshold pressure of ~11 Mpa, we can estimate the size of cavitation region $R_{cav}^{(num)} = 18\,\mu m$ at $V_0 = 20$kV. This size agrees with experimental observations, $R_{cav}^{(\exp)} = 15 - 17\,\mu m$. It is important to notice, that the electrostrictive negative pressure depends on the electrode tip radius as $p_E \propto E^2 \propto 1/R_0^2$, and therefore 30% uncertainty of the tip radius measurement leads to a change of threshold pressure value by two times.

Perturbation of liquid density $\delta\rho/\rho_0$ for different times for the case $V_0$=20 kV was presented in Figures 6 and 7. It is easy to see the density perturbation evolution. This explains the presence of the bright spots observed in Figure 4. The position of the maxima of luminosity in Fig. 4 coincides with the position of the maximum density of the liquid in Figure 6. Thus, at t = 88ns, in the experiment the maximum luminance is at position $l \approx 93\,\mu m$ from electrode (Figure 4), and the corresponding calculated maximum density at $l \approx 111\,\mu m$ (Fig. 6, line 5).

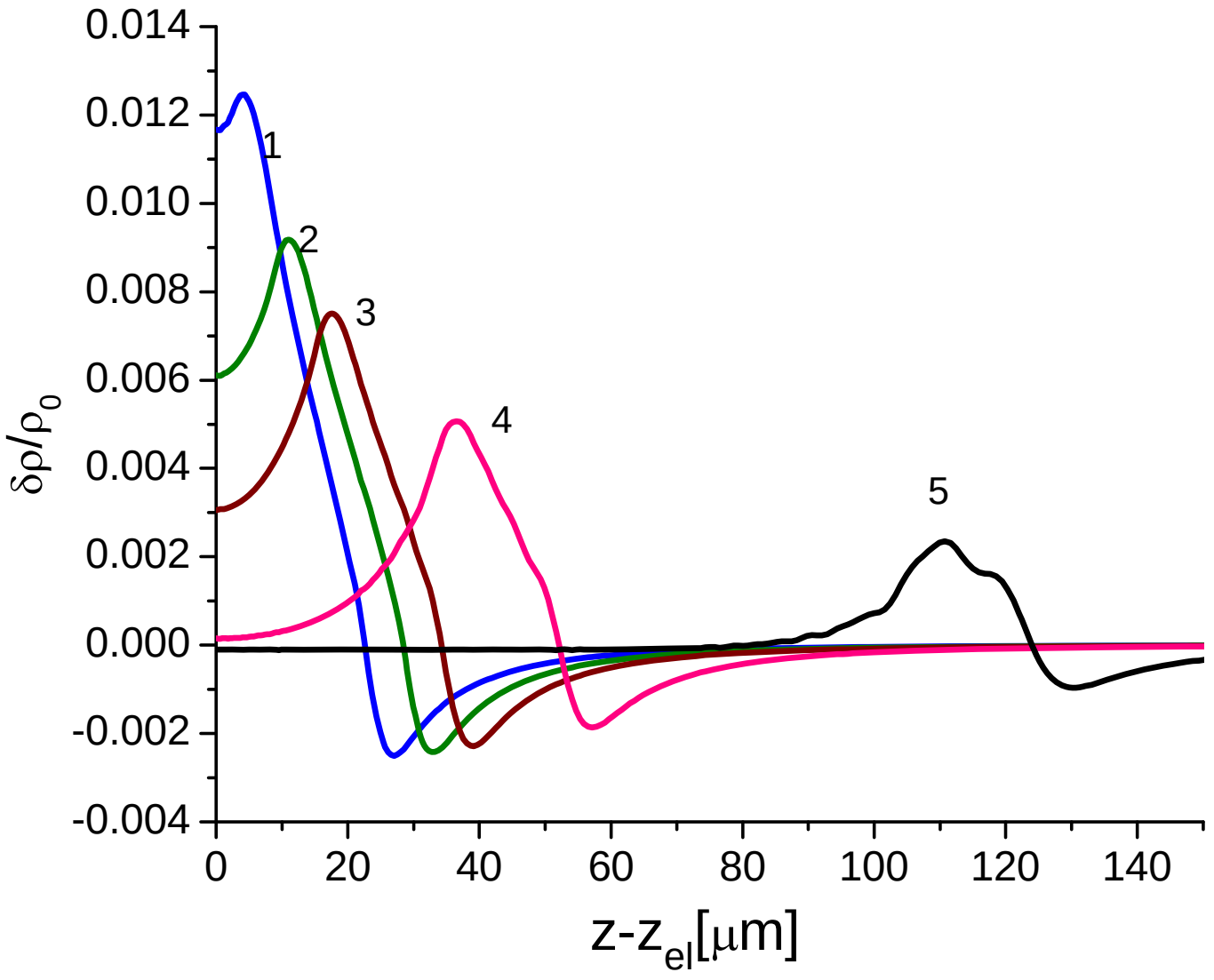

Fig. 6 Distributions of perturbation density $\delta n/n_0$ along axes of the electrode. Curve 1 corresponds to t=20 ns, 2 – 24 ns, 3 – 28 ns, 4 – 40ns, 5 – 88 ns.

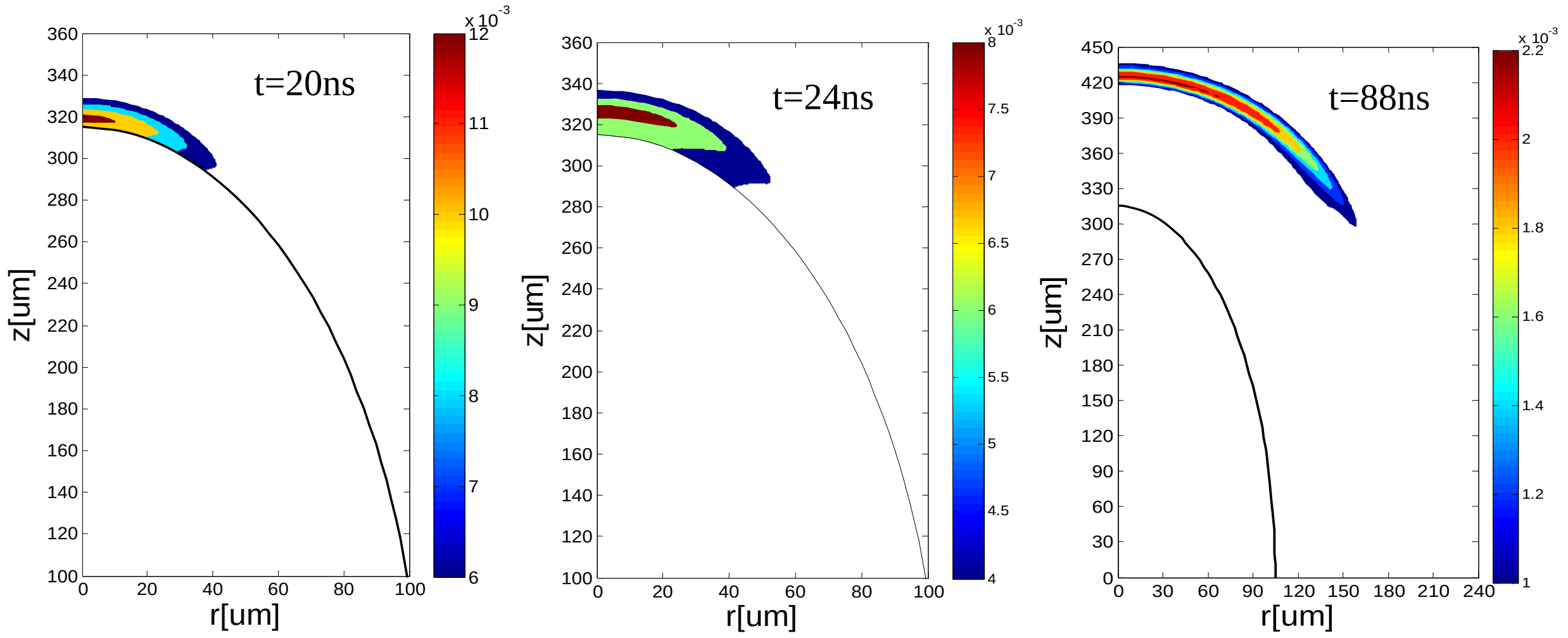

Fig. 7. Two dimensional distribution of perturbation density $\delta\rho/\rho_0$. Black line corresponds to electrode.

## Conclusions

It is shown that, in accordance with the theory, rapid pulsed voltage applied to the needle electrode, creates discontinuities in the fluid which can be detected by an optical method. There is good agreement between experiment and theory in all areas of measurements.